%% file: main.tex
\documentclass[5p]{elsarticle}
\biboptions{sort&compress}

\usepackage{amsmath}
\usepackage{lipsum}
\usepackage[dvipsnames]{xcolor}
\usepackage{subcaption}
\usepackage{graphicx}
\usepackage{siunitx}
\usepackage[colorlinks = true,linkcolor = blue,urlcolor  = blue, citecolor = blue]{hyperref}
\usepackage[normalem]{ulem}
\usepackage{amsmath,amsfonts,amssymb,amsthm} % AMS
\usepackage{verbatim }
\usepackage{tabularx}
\usepackage{cuted}
\usepackage{cleveref}
\usepackage{xurl}
\usepackage[flushleft]{threeparttable}
\usepackage{glossaries} 

\def\cm{c_\mathrm{m}}
\def\cti{c_{\mathrm{t},i}}

\def\cs{c_{\mathrm{s}}}
\def\Jvs{J_{\mathrm{vs}}}

\def\ksb{k_{\mathrm{sb}}}
\def\kbs{k_{\mathrm{bs}}}

\makeglossaries

\input{glossary}

\begin{document}

\begin{frontmatter}
\title{Kinetic surface model in FESTIM: Verification and Validation}
\cortext[cor]{Corresponding author}

\author[mephi]{Vladimir Kulagin\corref{cor}}
\ead{VVKulagin@mephi.ru}

\author[MIT]{Rémi Delaporte-Mathurin}
\author[CEA]{Etienne A. Hodille}
\author[IPP]{Mikhail Zibrov}

\address[mephi]{National Research Nuclear University MEPhI, Moscow, 115409, Russian Federation}
\address[MIT]{Plasma Science and Fusion Center, Massachusetts Institute of Technology, Cambridge, MA 02139, USA}
\address[CEA]{CEA, IRFM/GCFPM, F-13108 Saint-Paul-lez-Durance, France}
\address[IPP]{Max Planck Institute for Plasma Physics, Garching, 85748, Germany}

%\date{Submitted xx/xx/2023}
\journal{International Journal of Hydrogen Energy}

\begin{abstract}
The open-source code FESTIM (Finite Element Simulation of Tritium In Materials) is a powerful user-friendly software for hydrogen transport simulations. Up to now, it was capable of addressing various hydrogen transport problems with surface processes dependent on the concentration of solute species. However, the kinetics of surface hydrogen concentration should be considered under certain conditions. The recent 1.3 release of FESTIM introduced a new kinetic surface model, implemented in a flexible way for various applications. The correctness of the implementation is first proven using the method of manufactured solutions. Then, reliability of the model is demonstrated by reproducing four experimental cases on dynamics of hydrogen isotope retention in different materials. An additional cross-code comparison with two other simulation packages, MHIMS and TESSIM-X, shows an excellent agreement and strengthens the validity of the model.
\end{abstract}


\end{frontmatter}

\printglossary[nonumberlist, title=Nomenclature, style=symbunitlong]

\section{Introduction \label{sec:intro}}
Accumulation and transport of hydrogen (H) in materials are topical subjects of study in various research fields, such as nuclear fusion~\cite{Lu2014}, hydrogen storage and transport~\cite{Guo2023, Kobayashi2024}, material science~\cite{Li2020}, etc. Numerical modelling of H dynamics plays an important role in these fields allowing to perform prognostic estimations, validate theoretical models and construct new predictive ones. A number of dedicated computational packages exist (MHIMS~\cite{Hodille2024}, TESSIM-X~\cite{Schmid2016, Schmid2021}, TMAP7~\cite{osti_952013}, TMAP8~\cite{Lindsay2021}, FACE~\cite{Smirnov2017}, CRDS~\cite{Matveev2018}, etc.), which have their own capabilities and limitations. 

\begin{figure*}[ht]
    \centering 
    \includegraphics[scale=1]{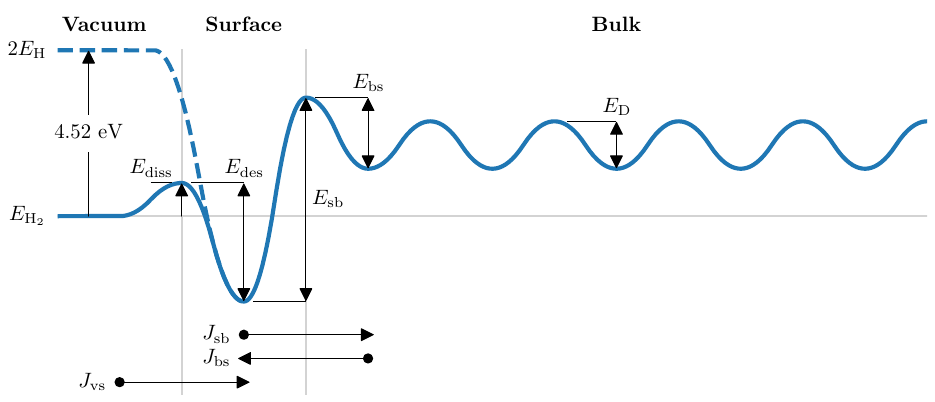}
    \caption{Near surface energy landscape of a hydrogen-metal system. Energy levels are measured from the H$_2$ state ($E_\mathrm{H_2}$). Arrows indicate H transition paths near the surface.}
    \label{fig:pot_diag}
\end{figure*}

Amongst the list of software, FESTIM (Finite Element Simulation of Tritium In Materials)~\cite{RDM2024_1} is one of a few open-source codes with an accessible Python interface and an active community. It is a flexible user-friendly package for the study of the H transport coupled with the heat transfer in 1D-3D multi-material geometries. In particular, FESTIM was successfully applied in fusion-related studies on hydrogen isotope retention in divertor monoblocks~\cite{RDM2021, RDM2024_1, Hodille2021}, tritium transport in a water-cooled lithium lead tritium breeding blanket~\cite{Dark2021}, tritium transport through nuclear fusion extractors~\cite{Dunnell2024}, effect of neutron irradiation on H retention~\cite{Dark2024}, impact of ELM-like exposure on H retention~\cite{Kulagin2025}, and other cases.

Up to now, the physics basis of FESTIM covered a wide range of bulk and surface processes, implemented dependent on the concentration of solute H. For the latter ones, this is a simplification relying on the use of effective rates and concentration of absorbed H near a surface. It is valid when surface processes are fast compared to bulk ones, so the concentrations of \textit{adsorbed} and \textit{absorbed} H are in equilibrium.

However, there are conditions~\cite{Schmid2021, Guterl2019, Pisarev1998}, when the distinction between concentrations of adsorbed and absorbed H should be made. A representative example is the H retention in a material exposed to a low-energy atomic/molecular flux. When the energy of incident particles is not high enough (less than several eVs) to overcome the surface barrier for implantation, incoming particles can stick to a surface and only then be absorbed into the bulk~\cite{Hoen2013,Maya2016,Ogorodnikova2016}. Another example is the surface kinetics of H at near-zero temperatures, when the probability of H absorption is low compared to other processes. Such conditions are usually met on dust grains in interstellar environment~\cite{Katz1999, Hama2013}.

The v1.3 release of FESTIM added a kinetic surface model, allowing users to distinguish between adsorbed and absorbed atoms and to simulate these surface processes explicitly. This feature opens opportunities for detailed studies of the H dynamics in materials during sorption experiments, transient events in tokamak plasma discharges, atomisation of H in capillary sources, etc. 

To ensure the correctness of the model and provide the reliability to the users, it has to be verified and validated (V\&V), in the same manner as it was done for other features of FESTIM~\cite{FESTIM_VV}. This paper describes the implementation of this new functionality. The main features of the model are provided in Section~\ref{sec:h_transport}. Section~\ref{sec:verification} presents the results of its verification using the method of manufactured solutions (MMS). The model validation against the experimental data is conducted in Section~\ref{sec:validation}. All input scripts used to obtain the results in this paper can be found at \href{https://github.com/KulaginVladimir/FESTIM-SurfaceKinetics-Validation}{github.com/KulaginVladimir/FESTIM-SurfaceKinetics-Validation}~\cite{SKVV_Git}.

\section{Hydrogen transport \label{sec:h_transport}}

\subsection{Bulk physics \label{subsec:bulk_physics}}
Hydrogen transport in the bulk is determined according to the model of McNabb \& Foster~\cite{McNabb1963}. The model considers two sorts of H species: mobile with concentration $\cm$ and trapped in defects of type $i$ with concentration $\cti$. 

\begin{figure*}[ht]
    \centering 
    \includegraphics[scale=1]{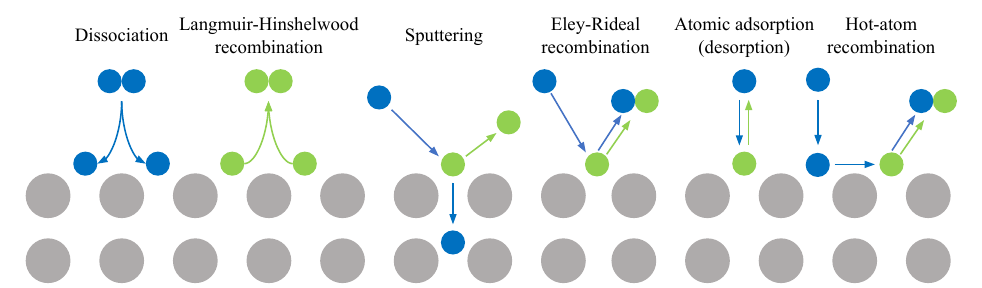}
    \caption{Possible surface processes taking place on a surface of a metal. Blue circles represent incident particles (atoms or molecules) from a near-surface environment and green circles - initially adsorbed atoms.}
    \label{fig:surf_processes}
\end{figure*}

In FESTIM, diffusion of mobile species can be defined by Fick's law (driven by a concentration gradient) and the Soret effect (driven by a temperature gradient). The latter is not considered in the present work, therefore the H diffusion flux $J$ is given by:
\begin{flalign}
    &\mathbf{J} = -D \nabla \cm&
\end{flalign}
where $D=D_0\exp\left(-E_\mathrm{D}/k_\mathrm{B} T\right)$ is the diffusion coefficient, $E_\mathrm{D}$ is the diffusion activation energy, $T$ is the absolute temperature.

Temporal evolution of mobile and trapped species is given by:
\begin{subequations}
\label{eq:bulk_conc}
\begin{flalign}
    &\frac{\partial \cm}{\partial t} = -\nabla \cdot \mathbf{J} + S - \sum\limits_i \frac{\partial \cti}{\partial t}
    \label{eq:mobile}&\\
    &\frac{\partial \cti}{\partial t}=k_i \, \cm \, (n_i - \cti)- p_i \, \cti&
    \label{eq:trapped}
\end{flalign}
\end{subequations}
where $S$ is the volumetric source of mobile H, $n_i$ is the concentration of traps $i$. The trapping and detrapping processes are assumed temperature-activated with the corresponding rates $k_i=k_{0,i}\exp\left(-E_{k,i}/k_\mathrm{B}T\right)$ and $p_i=p_{0,i}\exp\left(-E_{p,i}/k_\mathrm{B}T\right)$ expressed with the Arrhenius law.

\subsection{Kinetic surface model \label{subsec:kinetic_surface}}
The general approach to account for surface kinetics~\cite{Pick1985,Pisarev1997,Hodille2017,Schmid2021} consists in introducing a new concentration of hydrogen adsorbed on the surface $\cs$. The evolution of hydrogen surface concentration is determined by the atomic flux balance at the surface, as sketched in the simplified energy diagram (Fig.~\ref{fig:pot_diag}).

The governing equation for surface species is:
\begin{flalign}
    &\dfrac{d \cs}{d t} = J_{\mathrm{bs}} - J_{\mathrm{sb}} + \Jvs~\text{on}~\partial \Omega&    
    \label{eq:adsorbed}
\end{flalign}
where $J_\mathrm{bs}$ is the flux of hydrogen atoms from the subsurface (bulk region just beneath the surface) onto the surface, $J_\mathrm{sb}$ is the flux of hydrogen atoms from the surface into the subsurface, $\Jvs$ is the net flux of hydrogen atoms from the vacuum onto the surface, where $\partial\Omega$ is a boundary of a domain $\Omega$. It is worth noticing that the current implementation of the model does not account for possible surface diffusion and is applicable only for 1D hydrogen transport simulations. 

Compared to earlier codes, $\Jvs=\Jvs(T, \cm, \cs)$ is implemented as an arbitrary function and introduced to generalise the model for providing flexibility to the users. It is defined as $\Jvs=J_\mathrm{in}-J_\mathrm{out}$, where $J_\mathrm{in}$ is the sum of all fluxes coming from the vacuum onto the surface and $J_\mathrm{out}$ is the sum of all fluxes coming from the surface to the vacuum. $J_\mathrm{in}$ can be used to set up adsorption fluxes from different processes, for example, molecular dissociation or adsorption of low-energy atoms (see Fig.~\ref{fig:surf_processes}). Similarly, $J_\mathrm{out}$ can be used to define desorption fluxes governed by various processes, such as Langmuir-Hinshelwood recombination, Eley-Rideal recombination, sputtering, etc. 

The connection condition between surface and bulk domains represents the Robin boundary condition for the hydrogen transport problem:
\begin{flalign}
    &\mathbf{J} \cdot \mathbf{n} = \lambda_{\mathrm{IS}}\dfrac{\partial \cm}{\partial t} + J_{\mathrm{bs}} - J_{\mathrm{sb}}~\text{on}~\partial\Omega,&
    \label{eq:surfaceBC}
\end{flalign}
where $\lambda_\mathrm{IS}$ is the distance between two interstitial sites in the bulk. For a 1D problem, the normal vector \textbf{n} at the left boundary is $-\vec{x}$. The steady-state approximation of eq.~\eqref{eq:surfaceBC} at the left boundary is, therefore, $D\dfrac{\partial \cm}{\partial x}=J_\mathrm{bs}-J_\mathrm{sb}$ representing eq.~(12) in the original work of Pick \& Sonnenberg~\cite{Pick1985}.

The fluxes of subsurface-to-surface and surface-to-subsurface transitions are defined as follows~\cite{Hodille2017}:
\begin{subequations}
\begin{flalign}
    \label{eq:bs_sb_fluxes}
    &J_\mathrm{bs} = k_\mathrm{bs} \, \lambda_\mathrm{abs} \, c_\mathrm{m} \, \left(1-\dfrac{c_\mathrm{s}}{n_\mathrm{surf}}\right);&\\
    &J_\mathrm{sb} = k_\mathrm{sb} \, c_\mathrm{s} \, \left(1-\dfrac{\cm}{n_\mathrm{IS}}\right),&
\end{flalign}
\end{subequations}
where $n_\mathrm{surf}$ is the surface concentration of adsorption sites, $n_\mathrm{IS}$ is the bulk concentration of interstitial sites, $\lambda_\mathrm{abs}=n_\mathrm{surf}/n_\mathrm{IS}$ is the characteristic distance between surface and subsurface sites, $\kbs$ and $\ksb$ are the frequency factors for subsurface-to-surface and surface-to-subsurface transitions, respectively. The frequency factors can be set as functions of temperature, surface H concentration, subsurface H concentrations, as well as other user-defined parameters. 

These processes depend on the local concentration of H and are assumed to take place if there are available sites on the surface or in the subsurface. The terms in brackets account for possible surface/subsurface saturation. 

\section{Verification \label{sec:verification}}
Verification of the kinetic surface model is performed with the method of manufactured solutions (MMS). MMS is widely used to test the accuracy of numerical methods for solving systems of partial differential equations~\cite{Roache2002, Oberkampf_Roy_2010}. The approach consists in manufacturing an exact solution satisfying the system. Then, source terms and boundary conditions are obtained from the manufactured solution and provided in the numerical model. The error between the exact and computed solutions defines the accuracy of the numerical method.

The MMS test is a time-dependent H diffusion problem (eq.~\eqref{eq:mobile}) defined in the domain $\Omega:x\in[0,L]\cup t\in[0,t_\mathrm{f}]$, $L=1$, $t_\mathrm{f}=5$. Traps do not affect the surface processes and, therefore, are not considered. The chosen manufactured solution for mobile species is:
\begin{flalign}
    &c_\mathrm{m, exact}=1+2x^2+x+2t&
    \label{eq:cm_exact}
\end{flalign}
On the left boundary ($x=0$), the kinetic surface model (eq.~\eqref{eq:adsorbed}) is imposed, whereas on the right boundary ($x=L$) - a Dirichlet boundary condition is imposed: $\cm(x=1)=c_\mathrm{m, exact}(x=1)$. The diffusivity is $D=5$, the frequency factor of the subsurface-to-surface transition is $\kbs=1/\lambda_\mathrm{abs}$, the frequency factor of the surface-to-subsurface transition is $\ksb=2/\lambda_\mathrm{abs}$, the concentration of interstitial sites is $n_\mathrm{IS}=20$, the concentration of adsorption sites is $n_\mathrm{surf}=5$, and the characteristic distance between interstitial sites is $\lambda_\mathrm{IS}=2$. 

Injecting these parameters and the exact solution for the mobile concentration of H in eqs.~\eqref{eq:mobile} and \eqref{eq:adsorbed}, we obtain:
\begin{subequations}
\begin{flalign}
    & S = 2(1-2D)& \\
    & J_{\mathrm{vs}}=2n_\mathrm{surf}\dfrac{2n_\mathrm{IS}+2\lambda_\mathrm{IS}-D}{(2n_\mathrm{IS}-1-2t)^2}+2\lambda_\mathrm{IS}-D& \\
    & c_\mathrm{s, exact}=n_\mathrm{surf}\dfrac{1+2t+2\lambda_\mathrm{IS}-D}{2n_\mathrm{IS}-1-2t}&
\end{flalign}
\end{subequations}
The initial conditions for the simulation are $\cm(t=0)=c_\mathrm{m, exact}(t=0)$ and $\cs(t=0)=c_\mathrm{s, exact}(t=0)$. The simulation is conducted on the uniform mesh with 1000 elements and the fixed stepsize of \num{5e-3}. 

\begin{figure}[!t]
    \centering 
    \includegraphics[scale=1]{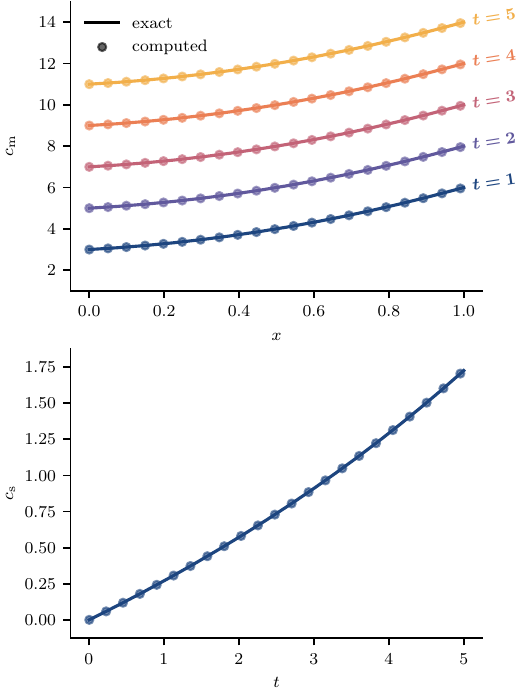}
    \caption{Evolution of the mobile concentration profile $\cm$ (top) and temporal evolution of the surface concentration $c_s$ (bottom).}
    \label{fig:MMS}
\end{figure}

To estimate the accuracy of the computed solution, the L2 errors are calculated:
\begin{subequations}
\begin{flalign}
    &E_\mathrm{bulk}=\sqrt{\int\limits_0^{t_\mathrm{f}} \int\limits_0^L \left(c_\mathrm{m, exact}-c_\mathrm{m, computed}\right)^2dxdt}&\\
    &E_\mathrm{surf}=\sqrt{\int\limits_0^{t_\mathrm{f}} \left(c_\mathrm{s, exact}-c_\mathrm{s, computed}\right)^2dt}&
\end{flalign}
\end{subequations}
The computed solutions show a good agreement with the exact ones (see Fig.~\ref{fig:MMS}). The L2 error for mobile H concentration is \num{2.33e-5} and the error for adsorbed H concentration is \num{4.29e-5}. These errors decrease with decreasing stepsizes.

\section{Validation \label{sec:validation}}
 This section presents four cases used to validate the implemented surface model. They reproduce experiments on the retention of hydrogen isotopes in Ti/W/EUROFER under exposure with low-energy atoms/molecules. In addition, three cases include a cross-code comparison between the results obtained with FESTIM and either MHIMS or TESSIM-X. For simulations, diffusivities of hydrogen isotopes in different materials were taken from the HTM database~\cite{HTM}, except the last validation case where we used the corresponding value from the original papers.
 
\subsection{Case 1: H absorption in Ti \label{subsec:case_1}}
The first validation case reproduces H absorption curves for Ti at different temperatures obtained by Hirooka et al.~\cite{Hirooka1981}. Absorption experiments were performed in the vacuum chamber at the base pressure of $P_\mathrm{H_2}=\SI{1.3e4}{Pa}$ with $10\times13\times1$ mm cold rolled Ti strips. Samples were kept at the constant temperature in the range from \SI{473}{K} to \SI{923}{K}. 

\begin{table}[!t]
    \caption{Input parameters for validation case 1~\cite{Shimohata2021}.}
    \label{tab:case1_inputs}
    \small
    \begin{threeparttable}
    \begin{tabularx}{\linewidth}{XXX}
         \hline
         Parameter & Unit & Value\\ 
         \hline
         $L$ & m & \num{1e-3} \\
         $n_\mathrm{Ti}$ & \si{m^{-3}} & \num{5.66e28}  \\ 
         $n_\mathrm{IS}$ & \si{m^{-3}} & $3n_\mathrm{Ti}$  \\ 
         $n_\mathrm{surf}$ & \si{m^{-2}} & \num{3.9e19} \\ 
         $\lambda_\mathrm{IS}$ & m & $n_\mathrm{surf}/n_\mathrm{IS}$\\
         $D_0$ & \si{m^2.s^{-1}} & \num{9e-7} \\
         $E_\mathrm{D}\tnote{*}$ & eV & 0.538 \\
         $E_\mathrm{des}$ & eV & 1.213 \\
         $E_\mathrm{bs}$ & eV & 0.881 \\
         $E_\mathrm{sb}$ & eV & 1.223 \\
         $E_\mathrm{diss} $ & eV & 0.0206 \\
         $\Phi_0$ &  & 0.0143\\
         $k_\mathrm{des,0}$ & \si{m^{2}.s^{-1}} & \num{4.271e-8} \\
         $k_\mathrm{bs,0}$ &  \si{s^{-1}} & \num{3.24e9} \\
         $k_\mathrm{sb,0}$ & \si{s^{-1}} & \num{1.16e11} \\
         $P_\mathrm{H_2,0}$ & Pa & \num{1.3e4} \\
         $V_\mathrm{ch}$ & \si{\meter^3} & \num{2.95e-3} \\
         \hline
    \end{tabularx}
    \begin{tablenotes}
    	\item[*] This is the value from the original work~\cite{Wille1981}, whereas Shimohata et al.~\cite{Shimohata2021} used a different value  
    \end{tablenotes}
    \end{threeparttable}
\end{table}

\begin{figure}[!t]
    \centering 
    \includegraphics[scale=1]{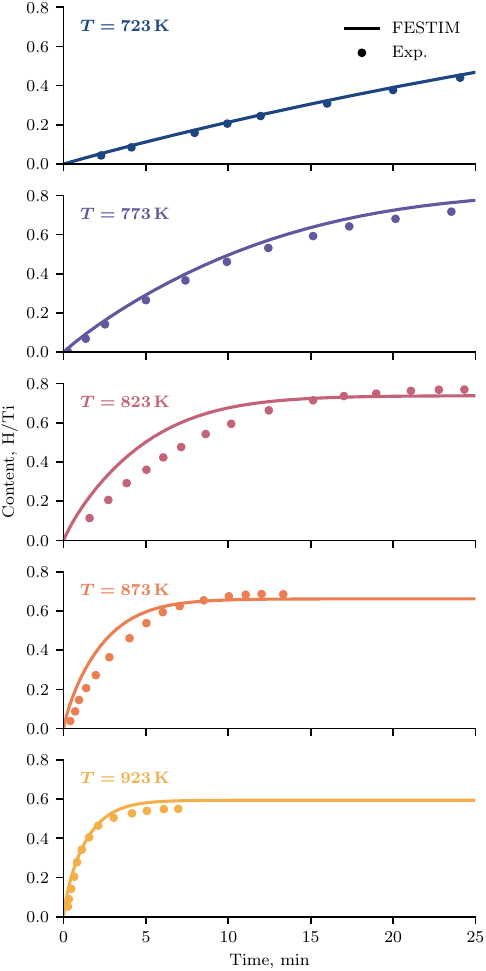}
    \caption{Temporal dependencies of the H content in Ti at different temperatures.}
    \label{fig:H_Ti}
\end{figure}

\begin{table*}[!t]
    \caption{Input parameters for validation case 2~\cite{Hodille2024}.}
    \label{tab:case2_inputs}
    \small
    \begin{tabularx}{\linewidth}{XXlXX}
         \cline{1-5}
         Parameter & Unit & \multicolumn{3}{l}{Value}\\ 
         \cline{3-5}
         && clean & 0.5ML of O & 0.75ML of O \\
         \cline{1-5}
         $L$ & m & \num{2e-3} \\
         $n_\mathrm{W}$ & m$^{-3}$ & \num{6.3382e28}  \\ 
         $n_\mathrm{IS}$ & m$^{-3}$ & $6n_W$  \\ 
         $\lambda_\mathrm{IS}$ & m & \num{1.117e-10} \\
         $\lambda_\mathrm{des}$ & m & $1/\sqrt{n_\mathrm{surf}}$ \\
         $\nu_0$ & s$^{-1}$ & \num{1e13} \\
         $D_0$ & m$^2$s$^{-1}$ & \num{1.93e-7}/$\sqrt{2}$  \\
         $E_\mathrm{D}$ & eV & 0.2\\
         $E_\mathrm{des}$ & eV & eq.~\eqref{eq:case2_Edes} \\
         $E_\mathrm{bs}$ &  eV & $E_\mathrm{D}$\\
         $E_\mathrm{sb}$  & eV & eq.~\eqref{eq:case2_Esb} \\
         $Q_\mathrm{S}$ & eV & 1 \\
         $E_\mathrm{diss} $ & eV & 0 \\
         $\Phi_0$ & & 1 \\
         $P_\mathrm{D_2}$ & Pa & \num{2e-5} \\
         $n_\mathrm{surf, clean}$ & m$^{-2}$  & \num{1.416e19} &  & \\ 
         $n_\mathrm{surf}$ & m$^{-2}$  & $n_\mathrm{surf,clean}$ & $0.5\,n_\mathrm{surf, clean}$ & $0.25\,n_\mathrm{surf, clean}$ \\ 
         $E_0$ & eV & 1.142 & 1.111 & 1.066 \\
         $\Delta E$ & eV & 0.346 & 0.289 & 0.234 \\
         $\theta_0$ & & 0.253 & 0.113 & 0.161 \\
         $\delta\theta$ & & 0.180 & 0.082 & 0.057 \\
         $\alpha$ & & 0.303 & 0.460 & 0.437 \\
         $\beta$ & & 8.902 & 7.240 & 4.144 \\
         \cline{1-5}
    \end{tabularx}
\end{table*}

The FESTIM model is based on the work of Shimohata et al.~\cite{Shimohata2021}. The evolution of the surface H concentration is assumed to be driven by adsorption from the gas phase ($J_\mathrm{ads}$) and recombination ($J_\mathrm{des}$). Therefore, the net flux of H atoms onto the surface from the vacuum in eq.~\eqref{eq:adsorbed} is:
\begin{subequations}
\label{eqs:case1_surf_proc}
\begin{flalign}
    &\Jvs=J_\mathrm{ads}-J_\mathrm{des}&\\
    &J_\mathrm{ads}=2 \, \Gamma_\mathrm{H_2} \, \Phi \, \left(1-\theta\right)^2&\\
    &J_\mathrm{des}=2 \, k_\mathrm{des} \, \cs^2&
\end{flalign}
\end{subequations}
where $\theta=\cs/n_\mathrm{surf}$ is the H surface coverage of Ti, $\Phi=\Phi_0 \exp(-E_\mathrm{diss}/k_\mathrm{B} T)$ is the sticking coefficient, $E_\mathrm{diss}$ is the activation energy for dissociation, $k_\mathrm{des}=k_\mathrm{des,0} \exp(-E_\mathrm{des}/k_\mathrm{B} T)$ is the desorption rate constant, and $E_\mathrm{des}$ is the desorption activation energy. The flux of molecules is:
\begin{subequations}
\label{eq:molecular_flux}
\begin{flalign}
    &\Gamma_\mathrm{H_2}=P_\mathrm{H_2}\, \Xi_\mathrm{H_2}&\\
    &\Xi_\mathrm{H_2}=\dfrac{1}{\sqrt{2\pi m_\mathrm{H_2}k_\mathrm{B} T}}&
\end{flalign}
\end{subequations}
where $\Xi_\mathrm{H_2}$ is the gas kinetic factor, $P_\mathrm{H_2}$ and $m_\mathrm{H_2}$ are the partial pressure and the mass of a H molecule, respectively. The chamber pressure is calculated with the ideal gas law:
\begin{flalign}
    &P_\mathrm{H_2} = \left( \frac{P_\mathrm{H_2,0} V_\mathrm{ch}}{k_\mathrm{B} T} + \frac{N_0-N}{2} \right) \frac{k_\mathrm{B} T}{V_\mathrm{ch}}&
\end{flalign}
where $P_\mathrm{H_2,0}$ is the initial pressure in the chamber, $V_\mathrm{ch}$ is the chamber volume, $N_0$ and $N$ are the initial and current amount of H in Ti, respectively.
The amount of H in Ti is calculated as:
\begin{flalign}
    &N=2A\left( \cs + \int\limits_0^L \cm dx \right)&
\end{flalign}
where $=\SI{1.3e-4}{m^2}$ is sample's cross-section area. The factor 2 is included, since only one half of the sample is simulated.  At each time step, the amount of H in Ti is evaluated and the chamber pressure is updated explicitly.

Temperature-dependent frequency factors for transitions between surface and subsurface are used:
\begin{subequations}
\label{eq:temperature_dependent_rates}
\begin{flalign}
    &\kbs = k_\mathrm{bs,0} \exp\left(-\frac{E_\mathrm{bs}}{k_\mathrm{B} T}\right)&\\
    &\ksb = k_\mathrm{sb,0} \exp\left(-\frac{E_\mathrm{sb}}{k_\mathrm{B} T}\right)&
\end{flalign}
\end{subequations}
Traps and sources of mobile H are not considered in the current case. The H diffusivity in Ti is taken from the report of Wille \& Davis~\cite{Wille1981}. The number of interstitial and surface sites are estimated via the bulk concentration of host atoms $n_\mathrm{Ti}$. Only a half of the sample is simulated with 1000-element uniform mesh and the kinetic surface model (eq.~\eqref{eq:surfaceBC}) imposed on the left boundary. The initial H content in Ti is set equal to zero. All input values for the mentioned parameters are listed in Table~\ref{tab:case1_inputs}. 

A reasonable agreement is achieved between the FESTIM results and the experimental data over the whole range of temperatures (see  Fig.~\ref{fig:H_Ti}). 

\subsection{Case 2: D adsorption on oxidised W \label{subsec:case_2}}
The second validation case reproduces experimental results on D desorption from oxidised W, performed by Dunand et al.~\cite{Dunand2022}. Single-crystalline W(110) samples (2 mm-thick) were prepared with different O coverages: clean, 0.5 monolayer (ML), and 0.75 ML. The samples were exposed to a D$_2$ gas at a base pressure of $P_\mathrm{D_2}=\SI{2e-5}{Pa}$ and at room temperature (\SI{300}{K}). The D$_2$ exposure lasted for 3000 s, followed by the storage phase for one hour. After this, thermal desorption spectroscopy (TDS) measurements were performed with a 5 K/s temperature ramp up to 800 K.

\begin{figure*}[ht]
    \centering 
    \includegraphics[scale=1]{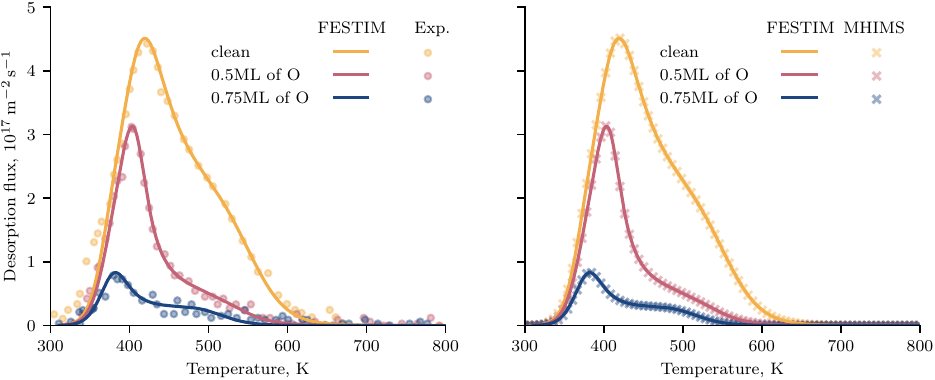}
    \caption{TDS spectra of D from W with different O coverages. Comparison with the experimental data (left) and MHIMS simulations (right).}
    \label{fig:D_WO}
\end{figure*}

Following the work of Hodille et al.~\cite{Hodille2024}, the evolution of the surface concentration $c_s$, in the present FESTIM model, is assumed to be governed by D adsorption from the gas phase and desorption (eqs.~\eqref{eqs:case1_surf_proc}). The flux of molecules $\Gamma_\mathrm{D_2}$ is calculated using eqs.~\eqref{eq:molecular_flux} with the mass of a D$_2$ molecule and partial pressure of D$_2$ in the chamber ($P_\mathrm{D_2}$). The presence of O on W affects the number of available adsorption sites for H:
\begin{flalign}
    &n_\mathrm{surf}=n_\mathrm{surf,clean}-n_\mathrm{surf,O}&
\end{flalign}
where $n_\mathrm{surf,clean}$ is the surface concentration of adsorption sites on a clean W(110) surface and $n_\mathrm{surf,O}$ is the surface concentration of O on a W surface.

\begin{figure*}
    \centering 
    \includegraphics[scale=1]{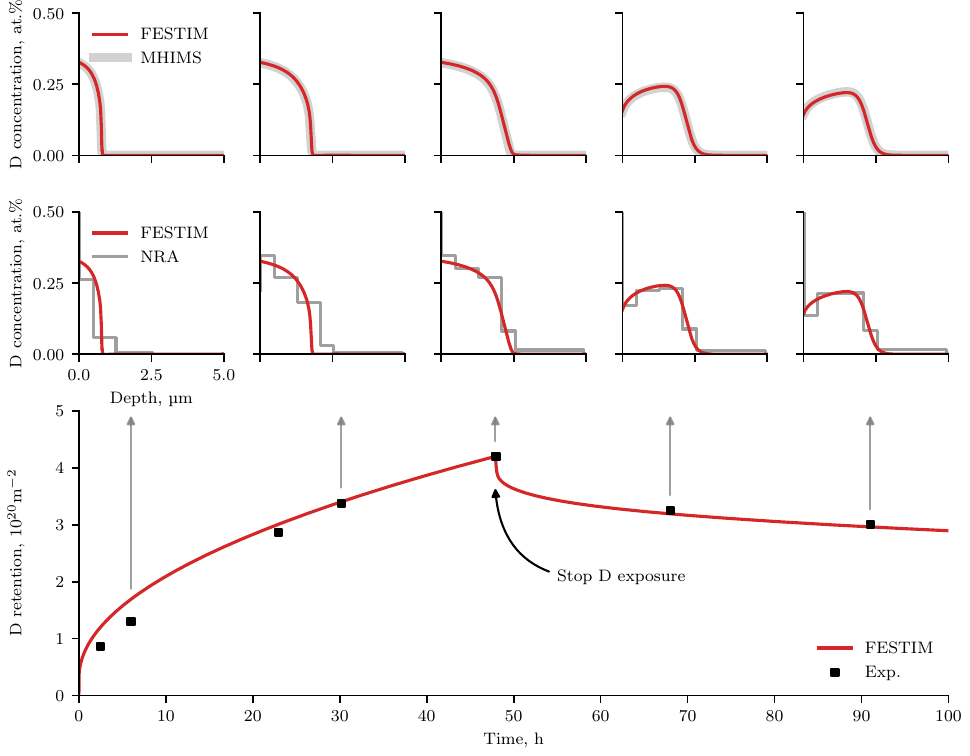}
    \caption{Temporal evolution of the D retention in W (bottom). Comparison between FESTIM and NRA measurements (middle) and FESTIM and MHIMS simulations (top).}
    \label{fig:D_damagedW}
\end{figure*}

According to the density functional theory calculations~\cite{Piazza2018, Ajmalghan2019, Bergstrom2019, Ferro2023}, the rate of desorption depends on the D coverage of W surface ($\theta=\cs/n_\mathrm{surf}$). To account for this effect, Hodille et al.~\cite{Hodille2024} used the coverage-dependent rate of desorption:
\begin{flalign}
    &k_\mathrm{des}=k_\mathrm{des,0} \exp \left( -\frac{E_\mathrm{des}(\theta)}{k_\mathrm{B} T} \right)&
\end{flalign}
where $k_\mathrm{des,0}=\nu_0\lambda_\mathrm{des}^2$, $\nu_0$ is the attempt frequency, and $\lambda_\mathrm{des}$ is the average distance between adsorption sites. The dependence of the activation energy for D desorption ($E_\mathrm{des}$) was approximated by the following expression:
\begin{flalign}
    &E_\mathrm{des}(\theta)=E_\mathrm{1}(\theta)\left( 1-\alpha\exp\left( -\beta(1-\theta) \right) \right)&\label{eq:case2_Edes}
\end{flalign}
$E_\mathrm{1}$ is expressed with a sigmoidal distribution and determines the change in the desorption energy below saturation:
\begin{flalign}
    &E_\mathrm{1}(\theta)=E_0 + \dfrac{\Delta E}{1+\exp\left( \dfrac{\theta-\theta_0}{\delta\theta} \right)}&
\end{flalign}
The second term in eq. (\ref{eq:case2_Edes}) is responsible for an exponential decrease of the desorption energy above the saturation~\cite{Matveev2018, Ferro2023}.

Temperature-dependent (eqs.~\eqref{eq:temperature_dependent_rates}) frequency factors of D transitions between surface and subsurface are used. Activation energy for re-absorption (i.e. surface-to-bulk transition) is chosen according to DFT calculations~\cite{Ajmalghan2019, Ferro2023}: $E_\mathrm{bs}\approx E_\mathrm{D}$, with the frequency pre-factor: $k_\mathrm{bs,0}=\nu_0$. The rate of D absorption depends on the surface coverage~\cite{Ajmalghan2019} and the material properties. Therefore, it is set as:
\begin{flalign}
    & \ksb = k_\mathrm{sb,0}\exp\left( -\frac{E_\mathrm{sb}(\theta)}{k_\mathrm{B} T} \right) & \\
    & E_\mathrm{sb}(\theta) = \frac{E_\mathrm{des}(\theta) - E_\mathrm{diss}}{2} + E_\mathrm{bs} + Q_\mathrm{S} \label{eq:case2_Esb} &
\end{flalign}
where $k_\mathrm{sb,0} = \nu_0$, $Q_\mathrm{S}=\SI{1}{eV}$~\cite{Fernandez2015} is the heat of D solution in W.

Since the absorption probability of D in W at room temperature is negligible, traps are not considered in the current case. However, the bulk parameters are required to fully pose the D transport problem in FESTIM. For modelling, the D diffusivity in W is defined by scaling the corresponding value for H~\cite{Fernandez2015} by a factor of $1/\sqrt{2}$. Similar to the previous case, the number of interstitial and surface sites is estimated via the bulk concentration of host atoms $n_\mathrm{W}$. A 500-element uniform mesh is used with the kinetic surface model imposed on the front surface ($x=0$) and a homogeneous Dirichlet boundary condition on the rear one ($\cm(x=L)=0$). At the start of simulations, the sample is assumed to be free of H. All input parameters are listed in Table~\ref{tab:case2_inputs}.

All TDS curves exhibit a low-temperature peak with a high-temperature shoulder (see Fig.~\ref{fig:D_WO}). With the increase in the O-coverage, the peak shifts to lower temperatures, and the overall amount of the released D decreases due to the reduction of the available adsorption sites on the surface. The FESTIM results are compared to the experimental data and the results of the MHIMS simulation, both taken from Hodille et al.~\cite{Hodille2024}. The simulated dependencies exhibit a good agreement with the experimental data and correlate perfectly with the MHIMS results.

\subsection{Case 3: D atom exposure of self-damaged W \label{subsec:case_3}}

The third validation case is NRA measurements of D in self-damaged W performed by Markelj et al.~\cite{Markelj2016}. The experimental procedure included three main phases. A polycrystalline W sample of 0.8 mm thickness was pre-damaged with 20 MeV W$^{6+}$ ions up to the fluence of \SI{7.8e17}{m^{-2}}. This pre-damaged sample was then exposed to a low-energy (\SI{\approx 0.3}{eV}) D flux of \SI{5.8e18}{m^{-2}.s^{-1}} at sample's temperature of \SI{600}{K}. The exposure continued until the D fluence of \SI{1e24}{m^{-2}} was reached. Finally, isothermal desorption of D for \SI{43}{h} at \SI{600}{K} was conducted.

\begin{table}[t]
    \caption{Input parameters for validation case 3~\cite{Hodille2015}.}
    \label{tab:case3_inputs}
    \small
    \begin{threeparttable}
    \begin{tabularx}{\linewidth}{XXX}
         \hline
         Parameter & Unit & Value\\ 
         \hline
         $L$ & m & \num{0.8e-3} \\
         $n_\mathrm{W}$ & \si{m^{-3}} & \num{6.3e28}  \\ 
         $n_\mathrm{IS}$ &\si{m^{-3}} & $6n_\mathrm{W}$  \\ 
         $n_\mathrm{surf}$ & \si{m^{-2}} & $6.9n_\mathrm{W}^{2/3}$ \\ 
         $\lambda_\mathrm{IS}$ & m & \num{110e-12}\\
         $\lambda_\mathrm{des}$ & m & $1/\sqrt{n_\mathrm{surf}}$\\
         $\nu_0$ & s$^{-1}$ & \num{1e13}\\
         $D_0$ & m$^2$s$^{-1}$ & \num{1.93e-7}/$\sqrt{2}$ \\
         $E_\mathrm{D}$ & eV & 0.2 \\
         $E_\mathrm{des}$ & eV & 1.74 \\
         $E_\mathrm{bs}$ & eV & $E_\mathrm{D}$ \\
         $E_\mathrm{sb}$ & eV & 1.545 \\
         $\Phi$ &  & 0.19\\
         $\Gamma_\mathrm{D}$ & \si{m^{-2}.s^{-1}} & \num{5.8e18} \\
         $\sigma_\mathrm{exc}$ & \si{m^{2}} & \num{1.7e-21} \\
         $n_1$ & m$^{-3}$ & $\num{1e-4}n_\mathrm{W}$  \\ 
         $n_2$ & m$^{-3}$ & $\num{1e-4}n_\mathrm{W}$  \\ 
         $n_3$ & m$^{-3}$ & $\num{1.9e-3}n_\mathrm{W}$  \\ 
         $n_4$ & m$^{-3}$ & $\num{1.6e-3}n_\mathrm{W}$  \\ 
         $n_5$ & m$^{-3}$ & $\num{2e-4}n_\mathrm{W}$  \\ 
         $E_\mathrm{p,1}$ & eV & 0.85  \\ 
         $E_\mathrm{p,2}$ & eV & 1.00  \\ 
         $E_\mathrm{p,3}$ & eV & 1.65  \\ 
         $E_\mathrm{p,4}$ & eV & 1.85  \\ 
         $E_\mathrm{p,5}$ & eV & 2.06  \\ 
         $p_\mathrm{0,1-5}$ & \si{s^{-1}} & $\nu_0$ \\
         $E_\mathrm{k,1-5}$ & eV & $E_\mathrm{D}$  \\ 
         $k_{0,1-5}$ & \si{m^{3}.s^{-1}} & $D_0/n_\mathrm{IS}\lambda_\mathrm{IS}^2$  \\
         \hline
    \end{tabularx}
    \end{threeparttable}
\end{table}

The FESTIM model is based on the approach of Hodille et al.~\cite{Hodille2017}. Only isothermal D exposure and desorption phases are simulated, omitting intermediate cooling/re-heating steps. Three surface processes are considered: adsorption of low-energy ions ($J_\mathrm{ads}$), desorption due to recombination of two adsorbed atoms ($J_\mathrm{des}$), and recombination of an incident atom with an adsorbed atom, i.e. Eley-Rideal recombination ($J_\mathrm{exc}$). Atomic fluxes due to the mentioned mechanisms are defined with:
\begin{subequations}
\label{eq:case3_Jvs}
\begin{flalign}
    &\Jvs=J_\mathrm{ads}-J_\mathrm{des}-J_\mathrm{exc}&  \\
    &J_\mathrm{ads}=\Gamma_\mathrm{D} \, \Phi \, \left( 1 - \theta \right)& \\
    &J_\mathrm{des}=2 \, k_\mathrm{des} \, \cs^2 & \\
    &J_\mathrm{exc}=\Gamma_\mathrm{D} \, \sigma_\mathrm{exc} \, \cs&
\end{flalign}    
\end{subequations}
where $\Gamma_\mathrm{D}$ is the flux of incident low-energy ions, $\sigma_\mathrm{exc}$ is the cross-section of the direct abstraction~\cite{Markelj2016}. The rate of desorption and frequency factors of subsurface-to-surface / surface-to-subsurface transitions are assumed to depend only on temperature. The pre-factors are: $k_\mathrm{des,0}=\nu_0 \lambda_\mathrm{des}^2$ and $k_\mathrm{bs,0}=k_\mathrm{sb,0}=\nu_0$.

For this problem, the same D diffusivity as in the previous validation case is used. Five types of trapping sites are included to reproduce the experimental data: two intrinsic traps, distributed homogeneously throughout the sample, and three extrinsic traps with a sigmoidal distribution ($f$) within the damaged layer:
\begin{flalign}
    &f(x)=\frac{1}{1+\exp\left( \dfrac{x-x_0}{\Delta x} \right)}&
\end{flalign}
where $x_0=\SI{2.2}{\micro\meter}$, $\Delta x=\SI{0.154}{\micro\meter}$.
Trapping in each defect site is assumed diffusion-limited. 

\begin{figure}[t]
    \centering 
    \includegraphics[scale=1]{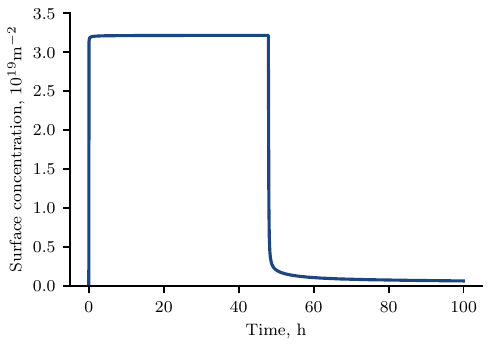}
    \caption{Temporal evolution of the D surface concentration.}
    \label{fig:surf_conc_evol}
\end{figure}

In the FESTIM model, we assumed that only the front surface ($x=0$) is exposed to D flux. Therefore, the full kinetic surface model (eqs.~\eqref{eq:case3_Jvs}) was imposed on the front surface, whereas only the desorption term in $\Jvs$ was accounted for on the opposite side ($x=L$). The simulations were conducted on a non-uniform mesh, including 800 elements. Zero initial conditions were used for concentrations of mobile and adsorbed H. Table~\ref{tab:case3_inputs} summarises the input values for this validation case.

During the exposure phase, the D retention increases (see the bottom plot in Fig.~\ref{fig:D_damagedW}) and the D concentration profile widens as deeper traps are being filled with D (see top/middle row of plots in Fig.~\ref{fig:D_damagedW}). After the exposure phase, a slower reduction in the D retention is observed due to the presence of traps with high ($>$\SI{1.5}{eV}) activation energy for detrapping. The results produced by FESTIM are in good agreement with the experimental data and correlate perfectly with MHIMS.

\begin{table}[t]
    \caption{Input parameters for validation case 4~\cite{Schmid2023_2}.}
    \label{tab:case4_inputs}
    \small
    \begin{threeparttable}
    \begin{tabularx}{\linewidth}{XXl}
         \hline
         Parameter & Unit & Value\\ 
         \hline
         $L$ & m & \num{0.8e-3} \\
         $n_\mathrm{EFe}$ & \si{m^{-3}} & \num{8.59e28}  \\ 
         $n_\mathrm{IS}$ &\si{m^{-3}} & $6n_\mathrm{EFe}$  \\ 
         $n_\mathrm{surf}$ & \si{m^{-2}} & $n_\mathrm{EFe}^{2/3}$ \\ 
         $\lambda_\mathrm{IS}$ & m & $n_\mathrm{EFe}^{-1/3}$ \\
         $D_0$ & m$^2$s$^{-1}$ & \num{1.5e-7} \\
         $E_\mathrm{D}$ & eV & 0.15 \\
         $E_\mathrm{des}$ & eV & 0.63 \\
         $E_\mathrm{diss}$ & eV & 0.4 \\
         $E_\mathrm{bs}$ & eV & $E_\mathrm{D}$ \\
         $Q_\mathrm{S}$\tnote{*} & eV & 0.238 \\
         $K_\mathrm{S,0}$\tnote{+} & \si{m^{-3}.Pa^{-1/2}} & \num{1.319e23} \\
         $\Phi_0$ &  & \num{1e-4} \\
         $\Gamma_\mathrm{D}$ & \si{m^{-2}.s^{-1}} & \num{5.8e18} \\
         $R_\mathrm{impl}$ & m & \num{-1e-10} \\
         $\sigma_\mathrm{impl}$ & m & \num{7.5e-10} \\
         $r$ & & 0.612 \\ 
         $\Omega_\mathrm{loss}$ &  & \num{8e4} \\
         $P_\mathrm{D_2}$ & \si{Pa} & 1 \\
         $k_\mathrm{des,0}$ & \si{\meter^2.s^{-1}} & \num{1e-4} \\
         $k_\mathrm{bs,0}$ & \si{s^{-1}} & $D_0/\lambda_\mathrm{IS}^2$ \\       
         $n_1$ & m$^{-3}$ & $\num{1e-5}n_\mathrm{EFe}$  \\ 
         $n_2$ & m$^{-3}$ & $2.5-5.0\times10^{-4}n_\mathrm{EFe}$  \\
         $E_\mathrm{p,1}$ & eV & 0.9  \\ 
         $E_\mathrm{p,2}$ & eV & 1.08  \\ 
         $p_\mathrm{0,1-2}$ & \si{s^{-1}} & \num{4e13} \\
         $E_\mathrm{k,1-2}$ & eV & $E_\mathrm{D}$  \\ 
         $k_{0,1-2}$ & \si{m^{3}.s^{-1}} & $D_0/n_\mathrm{IS}\lambda_\mathrm{IS}^2$  \\
         \hline
    \end{tabularx}
    \begin{tablenotes}
            \item[*] The value differs from that in the original work by Aiello et al.~\cite{Aiello2002} due to a conversion error in~\cite{Schmid2023_2}. For consistency, we use the value from~\cite{Schmid2023_2} in this comparison.
    	\item[+] The value differs from that in the original work by Aiello et al.~\cite{Aiello2002} due to a typographical error in~\cite{Aiello2002} itself.
    \end{tablenotes}
    \end{threeparttable}
\end{table}

An interesting finding in \cite{Hodille2015} is that the surface concentration of D evolves much faster than the bulk concentration. The surface is covered with D within several minutes (see Fig.~\ref{fig:surf_conc_evol}). Then, the surface concentration reaches the steady-state regime. Finally, the major amount of H is being rapidly released from the surface during the desorption phase, and the surface concentration approaches the next plateau. This result allows to consider a steady-state approximation of the kinetic surface model, which can be more computationally efficient for simulating the retention of hydrogen isotopes on large time-scales, relevant to plasma discharges in fusion devices. 

\subsection{Case 4: D implantation in W-damaged EUROFER \label{subsec:case_4}}

The last validation cases is dedicated to D retention in displacement-damaged EUROFER under low-energy ion exposure, studied by Schmid et al.~\cite{Schmid2023_1}. Experiments were conducted with three types of EUROFER samples (thickness of 0.8 mm): undamaged, damaged with 20 MeV W$^{6+}$ ions, pre-loaded with D and then damaged with 20 MeV W$^{6+}$ ions. These samples were then irradiated with a low-energy (5 eV/ion) D flux of $\approx\SI{9e19}{\meter^{-2}s^{-1}}$ at the gas pressure of $\SI{1}{Pa}$ and at temperature of \SI{370}{K}. The exposure time varied in these experiments between \SI{48}{h} and \SI{143}{h} resulting in four cases:
\begin{itemize}
    \item \SI{143}{h} plasma: undamaged sample loaded for 143 h with D;
    \item DPA$\rightarrow$\SI{48}{h} plasma: damaged sample loaded for \SI{48}{h} with D;
    \item DPA$\rightarrow$\SI{143}{h} plasma: damaged sample loaded for \SI{143}{h} with D;
    \item DPA+D$\rightarrow$\SI{48}{h} plasma: pre-loaded damaged sample exposed for \SI{48}{h} with D.
\end{itemize}
After exposure, the samples were stored for \SI{24}{h} at \SI{290}{K}. Finally, TDS measurements of D desorption were performed with the temperature ramp of $\approx\SI{3}{K/min}$ up to \SI{800}{K}.

The FESTIM model is based on the work of Schmid et al.~\cite{Schmid2023_2}. Similar to the previous validation case, three mechanisms are included in the kinetic surface model: dissociation of D$_2$ molecules ($J_\mathrm{ads}$), recombination of D$_2$ molecules ($J_\mathrm{des}$), and an abstraction flux ($J_\mathrm{loss}$). The last process is similar to the Eley-Rideal recombination, but was introduced by the authors as an \textit{ad hoc} loss channel of adsorbed D due to impact of energetic ions. The net flux of atoms to the surface is:
\begin{subequations}
\label{eq:case4_Jvs}
\begin{flalign}
    &\Jvs=J_\mathrm{ads}-J_\mathrm{des}-J_\mathrm{loss}&  \\
    &J_\mathrm{ads}=2 \, \Gamma_\mathrm{D_2} \, \Phi \, \left( 1 - \theta \right)^2& \\
    &J_\mathrm{des}=2 \, k_\mathrm{des} \, \cs^2 & \\
    &J_\mathrm{loss}=\Gamma^\mathrm{impl}_\mathrm{D} \, \Omega_\mathrm{loss} \, \theta&
\end{flalign}    
\end{subequations}
where $\Omega_\mathrm{loss}$ can be treated as an effective yield of D removal due to incoming flux and $\Gamma^\mathrm{impl}_\mathrm{D}=\Gamma_\mathrm{D}\,(1-r)$ is the flux of implanted D atoms, $r$ is the reflection coefficient. The flux of molecules ($\Gamma_\mathrm{D_2}$) is calculated using eqs.~\eqref{eq:molecular_flux}.

The subsurface-to-surface transition is assumed diffusion-limited: $E_\mathrm{bs}=E_\mathrm{D}$ and $k_\mathrm{bs,0}=D_0/\lambda_\mathrm{IS}^2$, where $\lambda_\mathrm{IS}=\lambda_\mathrm{abs}=n_\mathrm{EFe}^{-1/3}$ is the typical lattice spacing in EUROFER and $n_\mathrm{EFe}$ is the concentration of host atoms in EUROFER. The frequency factor of surface-to-subsurface transition is chosen so Sieverts' law is satisfied:
\begin{flalign}
    \label{eq:case4_ksb}
    &\ksb=\kbs \, K_\mathrm{S} \, \lambda_\mathrm{IS} \, \sqrt{\frac{k_\mathrm{des}}{\Phi\Xi_\mathrm{D_2}}}&
\end{flalign}
where $K_\mathrm{S}=K_\mathrm{S,0}\exp(-Q_\mathrm{S}/k_\mathrm{B}T)$ is the solubility.

\begin{figure}[!t]
    \centering 
    \includegraphics[scale=1]{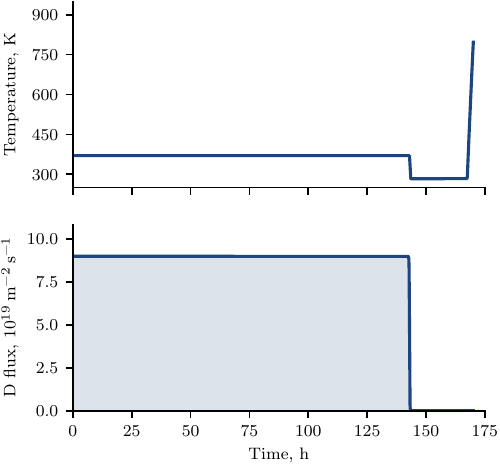}
    \caption{Temporal evolutions of material temperature (top) and D flux (bottom) used in "\SI{143}{h} plasma" simulation case.}
    \label{fig:TFlux_evol}
\end{figure}

\begin{figure*}[t]
    \centering 
    \includegraphics[scale=1]{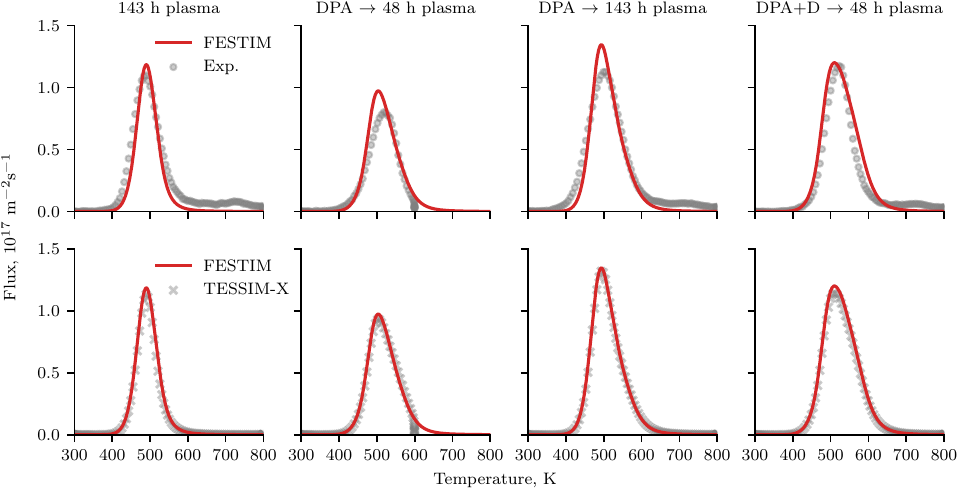}
    \caption{TDS spectra of D from different EUROFER samples. Comparison with the experimental data (top) and TESSIM-X
simulations (bottom).}
    \label{fig:D_eurofer}
\end{figure*}

The D diffusivity in EUROFER is defined with the fitted parameters from the original work of Aiello et al.~\cite{Aiello2002}. However, it must be noted that these parameters were determined for a limited temperature range (623-\SI{722}{K}), which does not cover the one considered in this validation case. In all cases, intrinsic trapping sites ($n_1$) are considered with a homogeneous distribution within the sample. For damaged samples, additional extrinsic traps ($n_2$) are included with the following distribution within the damaged zone:
\begin{flalign}
    &f(x)=0.5 \, \left( 1-\tanh \left( \frac{x-x_0}{\Delta x} \right)\right)&
\end{flalign}
where $x_0=\SI{3.3}{\micro\meter}$ and $\Delta x = \SI{0.01}{\micro\meter}$. The concentration of extrinsic traps varied depending on the presence of D during damaging. It was measured to be $\num{2.5e-4}n_\mathrm{EFe}$ in the case without D exposure during damaging and $\num{5.0e-4}n_\mathrm{EFe}$ otherwise. Trapping in each defect type is assumed diffusion-limited. 

The front surface (at $x=0$) is assumed to be irradiated with the flux of energetic D ions. Therefore, the full kinetic surface model is imposed on the front surface. Moreover, an implantation source of mobile D ($S$ term in eq.~\eqref{eq:mobile}) near the exposed surface is included with a normalised Gaussian profile, defined with the fitted mean range ($R_\mathrm{impl}$) and standard deviation ($\sigma_\mathrm{impl}$) from~\cite{Schmid2023_2}. On the rear boundary ($x=L$), the dissociation flux and loss flux of adsorbed D due to the incidence of energetic ions ($J_\mathrm{loss}$) are not accounted. 

To reproduce the experimental data and perform a cross-code comparison, we used the same smooth expressions for temperature and D flux evolutions as for the TESSIM-X simulations (see example in Fig.~\ref{fig:TFlux_evol}).

The simulations were conducted on a non-uniform mesh consisting of 650 elements with zero initial conditions for both mobile and adsorbed D. The list of input values is shown in Table~\ref{tab:case4_inputs}. Certain required parameters not specified in \cite{Schmid2023_2} were obtained through private communication. Additionally, some parameters differ between the published paper and the input files of TESSIM-X.

Existence of displacement-damage traps in EUROFER slightly shifts the peaks of TDS profiles to higher temperature region (see Fig.~\ref{fig:D_eurofer}. It also widens the profiles having more gradually decreasing high-temperature shoulders. Although the agreement between FESTIM results and experimental data (top row in Fig.~\ref{fig:D_eurofer}) is satisfactory, the correlation with TESSIM-X is excellent (bottom row in Fig.~\ref{fig:D_eurofer}). Minor discrepancies can be attributed to differences in the implementation of the kinetic surface model as well as input parameters.  

\section{Conclusion \label{sec:conclusion}}
The recent release of FESTIM v1.3 introduced a new model allowing to account for kinetics of H on material surfaces. The implemented model has been verified and validated, demonstrating its correctness and applicability for the study of various H transport problems. Comparison of FESTIM results with MHIMS and TESSIM-X codes has shown a remarkable agreement, providing additional reliability to users. In addition, the described simulation cases will be included in the FESTIM V\&V book.

There are plans for further development of the kinetic surface model. The current implementation supports only 1D H transport simulations. Extending the model to a 2D/3D dimensional case will make it possible to analyse the effect of surface diffusion on the H retention dynamics. 

Another promising modification is the steady-state approximation of surface kinetics. We have shown there are conditions when the concentration of adsorbed H equilibrates rapidly. This steady-state model can reduce the efforts needed to simulate the dynamics of H on large time-scales, while considering the surface physics. Finally, the model is planned to be ported to the upcoming major release of FESTIM 2, which utilises an up-to-date framework with even wider functionality.  

\section*{CRediT author statement}
\textbf{Vladimir Kulagin}: 
Conceptualisation, Software, Visualisation, Writing – original draft.

\textbf{Rémi Delaporte-Mathurin}:
Conceptualisation, Software, Visualisation, Writing — review \& editing, Supervision.

\textbf{Etienne A. Hodille}:
Data Curation, Writing — review \& editing.

\textbf{Mikhail Zibrov}:
Data Curation, Writing — review \& editing.

% \section*{Acknowledgements}

\bibliography{References.bib}
\end{document}

%% file: glossary.tex
\glsaddkey{unit}{\glsentrytext{\glslabel}}{\glsentryunit}{\GLsentryunit}{\glsunit}{\Glsunit}{\GLSunit}

\newglossarystyle{symbunitlong}{%
\setglossarystyle{long3col}% base this style on the list style
  {\end{supertabular}}%

}

\newglossaryentry{D}
{
    name=$D$,
    description={Diffusion coefficient},
    unit={\si{m^2.s^{-1}}},
}

\newglossaryentry{mobile_conc}
{
    name=$\cm$,
    description={Mobile concentration},
    unit={\si{m^{-3}}},
}    

\newglossaryentry{trapped_conc}
{
    name=$\cti$,
    description={Particle concentration in the $i$-th trap},
    unit={\si{m^{-3}}}
} 

\newglossaryentry{surf_conc}
{
    name=$\cs$,
    description={Particle concentration on a surface},
    unit={\si{m^{-2}}}
} 

\newglossaryentry{diffusion_flux}
{
    name=$J$,
    description={Diffusion flux},
    unit={\si{m^{-2}.s^{-1}}}
}

\newglossaryentry{particle_flux_vs}
{
    name=$\Jvs$,
    description={Net particle flux from the vacuum onto the surface},
    unit={\si{m^{-2}.s^{-1}}}
}

\newglossaryentry{E_sb}
{
    name=$E_{\mathrm{sb}}$,
    description={Activation energy for surface-to-subsurface transition},
    unit={\si{eV}}
}

\newglossaryentry{E_bs}
{
    name=$E_{\mathrm{bs}}$,
    description={Activation energy for subsurface-to-surface transition},
    unit={\si{eV}}
}

\newglossaryentry{E_des}
{
    name=$E_{\mathrm{des}}$,
    description={Activation energy for desorption},
    unit={\si{eV}}
}

\newglossaryentry{E_D}
{
    name=$E_{\mathrm{D}}$,
    description={Diffusion activation energy},
    unit={\si{eV}}
}

\newglossaryentry{E_k}
{
    name=$E_{k,i}$,
    description={Activation energy for H trapping in the $i$-th trap},
    unit={\si{eV}}
}

\newglossaryentry{E_p}
{
    name=$E_{p,i}$,
    description={Activation energy for H detrapping from the $i$-th trap},
    unit={\si{eV}}
}

\newglossaryentry{flux_sb}
{
    name=$J_{\mathrm{sb}}$,
    description={Particle flux from the surface into the subsurface},
    unit={\si{m^{-2}.s^{-1}}}
}

\newglossaryentry{flux_bs}
{
    name=$J_{\mathrm{bs}}$,
    description={Particle flux from the subsurface onto the surface},
    unit={\si{m^{-2}.s^{-1}}}
}

\newglossaryentry{transition_rate_des}
{
    name=$k_\mathrm{des}$,
    description={Desorption rate constant},
    unit={\si{m^{2}.s^{-1}}}
}

\newglossaryentry{transition_rate_sb}
{
    name=$\ksb$,
    description={Frequency factor of the surface-to-subsurface transition},
    unit={\si{s^{-1}}}
}

\newglossaryentry{transition_rate_bs}
{
    name=$\kbs$,
    description={Frequency factor of the subsurface-to-surface transition},
    unit={\si{s^{-1}}}
}

\newglossaryentry{particle_source}
{
    name=$S$,
    description={Particle source term},
    unit={\si{m^{-3}.s^{-1}}}
}

\newglossaryentry{time}
{
    name=$t$,
    description={Time},
    unit={\si{s}}
}

\newglossaryentry{trapping_rate}
{
    name=$k_i$,
    description={Trapping rate in $i$-th tapping site},
    unit={\si{m^3.s^{-1}}}
}

\newglossaryentry{detrapping_rate}
{
    name=$p_i$,
    description={Detrapping rate from $i$-th trapping site},
    unit={\si{s^{-1}}}
}

\newglossaryentry{trap_dens}
{
    name=$n_i$,
    description={Concentration of $i$-th trapping sites},
    unit={\si{m^{-3}}}
}

\newglossaryentry{pressure}
{
    name=$P$,
    description={Pressure},
    unit={\si{Pa}}
}

\newglossaryentry{temp}
{
    name=$T$,
    description={Temperature},
    unit={\si{K}}
}

\newglossaryentry{normal}
{
    name=$\textbf{n}$,
    description={Normal vector},
    unit={}
}

\newglossaryentry{l_IS}
{
    name=$\lambda_\mathrm{IS}$,
    description={Distance between interstitial sites},
    unit={\si{m}}
}

\newglossaryentry{l_abs}
{
    name=$\lambda_\mathrm{abs}$,
    description={Distance between adsorption and interstitial sites},
    unit={\si{m}}
}

\newglossaryentry{l_des}
{
    name=$\lambda_\mathrm{des}$,
    description={Distance between adsorption sites},
    unit={\si{m}}
}

\newglossaryentry{n_IS}
{
    name=$n_\mathrm{IS}$,
    description={Concentration of interstitial sites},
    unit={\si{m^{-3}}}
}

\newglossaryentry{n_X}
{
    name=$n_\mathrm{X}$,
    description={Concentration of host atoms in material X},
    unit={\si{m^{-3}}}
}

\newglossaryentry{n_surf}
{
    name=$n_\mathrm{surf}$,
    description={Concentration of adsorption sites},
    unit={\si{m^{-2}}}
}

\newglossaryentry{Phi}
{
    name=$\Phi$,
    description={Sticking coefficient},
    unit={}
}

\newglossaryentry{E_diss}
{
    name=$E_\mathrm{diss}$,
    description={Activation energy for dissociation},
    unit={eV}
}

\newglossaryentry{heat_solution}
{
    name=$Q_\mathrm{S}$,
    description={Heat of solution},
    unit={eV}
}

\newglossaryentry{nu}
{
    name=$\nu_0$,
    description={Attempt frequency},
    unit={\si{s^{-1}}}
}

\newglossaryentry{thickness}
{
    name=$L$,
    description={Geometry thickness},
    unit={\si{m}}
}

\newglossaryentry{r_refl}
{
    name=$r$,
    description={Reflection coefficient},
    unit={}
}

\newglossaryentry{X_impl}
{
    name=$R_\mathrm{impl}$,
    description={Mean implantation range of mobile atoms},
    unit={\si{m}}
}

\newglossaryentry{sigma_impl}
{
    name=$\sigma_\mathrm{impl}$,
    description={Standard deviation of implantation profile of mobile atoms},
    unit={\si{m}}
}

\newglossaryentry{Boltzmann_const}
{
    name=$k_\mathrm{B}$,
    description={The Boltzmann constant},
    unit={\si{ev/K}}
}

\newglossaryentry{trap_distr}
{
    name=$f$,
    description={Distribution of extrinsic traps},
    unit={}
}

\newglossaryentry{sigma_exc}
{
    name=$\sigma_\mathrm{exc}$,
    description={Cross-section of recombination of an incident D atom with an adsorbed D atom on a surface},
    unit={\si{m^2}}
}

\newglossaryentry{omega_loss}
{
    name=$\Omega_\mathrm{loss}$,
    description={Effective yield of D removal due to incoming flux},
    unit={}
}

\newglossaryentry{molecule_mass}
{
    name=$m_\mathrm{X}$,
    description={Mass of molecule of hydrogen isotope X},
    unit={\si{kg}}
}

\newglossaryentry{surface_coverage}
{
    name=$\theta$,
    description={Surface coverage},
    unit={}
}
\glsaddallunused